\documentclass[aip,reprint]{revtex4-1}

\draft 

\usepackage[dvipdfmx]{graphicx}
\usepackage{color}
\usepackage{bm}
\usepackage{amsmath,amssymb}
\usepackage{braket}

\begin{document}


\title[Single-photon generation from a neodymium ion in optical fiber at room temperature]{Single-photon generation from a neodymium ion in optical fiber at room temperature} 



\author{Kaito Shimizu}
\email{1223702@ed.tus.ac.jp}
\affiliation{Depertment of Physics, Tokyo University of Science, Kagurazaka 1-3, Shinjuku, Tokyo, 162-8601, Japan}

\author{Kai Inoue}
\affiliation{Depertment of Physics, Tokyo University of Science, Kagurazaka 1-3, Shinjuku, Tokyo, 162-8601, Japan}

\author{Kazutaka Katsumata}
\affiliation{Depertment of Physics, Tokyo University of Science, Kagurazaka 1-3, Shinjuku, Tokyo, 162-8601, Japan}

\author{Ayumu Naruki}
\affiliation{Depertment of Physics, Tokyo University of Science, Kagurazaka 1-3, Shinjuku, Tokyo, 162-8601, Japan}

\author{Mark Sadgrove}
\affiliation{Depertment of Physics, Tokyo University of Science, Kagurazaka 1-3, Shinjuku, Tokyo, 162-8601, Japan}

\author{Kaoru Sanaka}
\affiliation{Depertment of Physics, Tokyo University of Science, Kagurazaka 1-3, Shinjuku, Tokyo, 162-8601, Japan}



\date{\today}

\begin{abstract}
The realization of single-photon generation is important for implementing various quantum information technologies. The use of rare-earth ions in an optical fiber is a promising single photon generation method due to its ability to operate at room temperature as well as the low cost involved. Neodymium ions are especially interesting because the ions are one of the most commercially affordable rare-earth materials in the current industry. The neodymium ion also has the advantage of having a rich energy level structure, which offers several possible wavelengths for emitted single photons from visible to near-telecommunication wavelengths. In this paper, we experimentally demonstrated single-photon generation using an isolated single neodymium ion in tapered silica fiber at room temperature. Our results have significant implications as a platform for low-cost wavelength-selectable single-photon sources and photonic quantum applications.
\end{abstract}


\maketitle 

Single-photon sources are attracting much attention as a component of various quantum information technologies, such as quantum key distribution~\cite{qkd01, qkd02}, quantum random number generation~\cite{random01, random02}, and quantum computing~\cite{computer01}. Thus, several methods of single-photon generation using various kinds of quantum emitters, such as single atoms~\cite{atom01}, single ions~\cite{ion01}, quantum dots~\cite{qdot01, qdot02}, nitrogen vacancy centers in solid-state materials~\cite{nvcenter01, nvcenter02}, silicon-vacancy centers in nanodiamonds~\cite{sv01, sv02}, and hexagonal boron nitrides~\cite{hbn01, hbn02}, have been proposed to realize single-photon sources for applications in such quantum technologies. Among them, rare-earth (RE) atoms are among the best candidates as a single-photon source because of their stability in emitting photons with the transition of 4f orbital electrons, simplicity of operation at room temperature, and compatibility with optical fiber for establishing long-distance quantum communication technologies~\cite{fibercom01}.

Furthermore, RE atoms have a long coherence time because they have milli-second level long fluorescence lifetimes. This feature allows quantum entangled states to be distributed over long distances and is significant for the realization of long-range quantum networks~\cite{rare-earth02}. However, this also leads to low brightness, although this can be overcome by the Purcell effect, which occurs by combining resonators~\cite{Purcell01}.

Recently, we proposed a method of single-site optical spectroscopy and optically addressed single RE atoms doped in an amorphous silica optical fiber at room temperature by tapering RE atoms doped in optical fiber, and we reported the experimental results using ytterbium ion(Yb$^{3+}$)-doped optical fiber~\cite{Yb01}.
In this paper, we demonstrate similar experimental results using a neodymium ion(Nd$^{3+}$)-doped optical fiber at room temperature.
We also studied the dependence of polarization states of photons emitted from a Nd$^{3+}$ ion in the optical fiber.
In addition, we experimentally measured the saturation intensity of a pump laser for a single Nd$^{3+}$ ion in tapered fiber and theoretically analyzed the saturation intensity of the two-level system in the fiber.

The fiber-coupled single-photon sources have been studied and still under investigation. In conventional methods, the fiber-coupled single-photon source is given by combining the tapered fiber with a single photon emitter from outside of the tapered fiber~\cite{fibercoupled01, fibercoupled02}. Using this method, it is hard to experimentally demonstrate high channeling efficiency in the direction of the guided mode of the fiber. In addition, the channeling efficiency is about up to 30\,\% at theoretical value~\cite{fibercoupled03}.
In our method, the channeling efficiency can reach about 32\,\% without the use of special techniques to improve because we used initially RE doped fiber into the fiber~\cite{Yb01}.

 \begin{figure}[htp]
 \centering
\includegraphics[width=8cm]{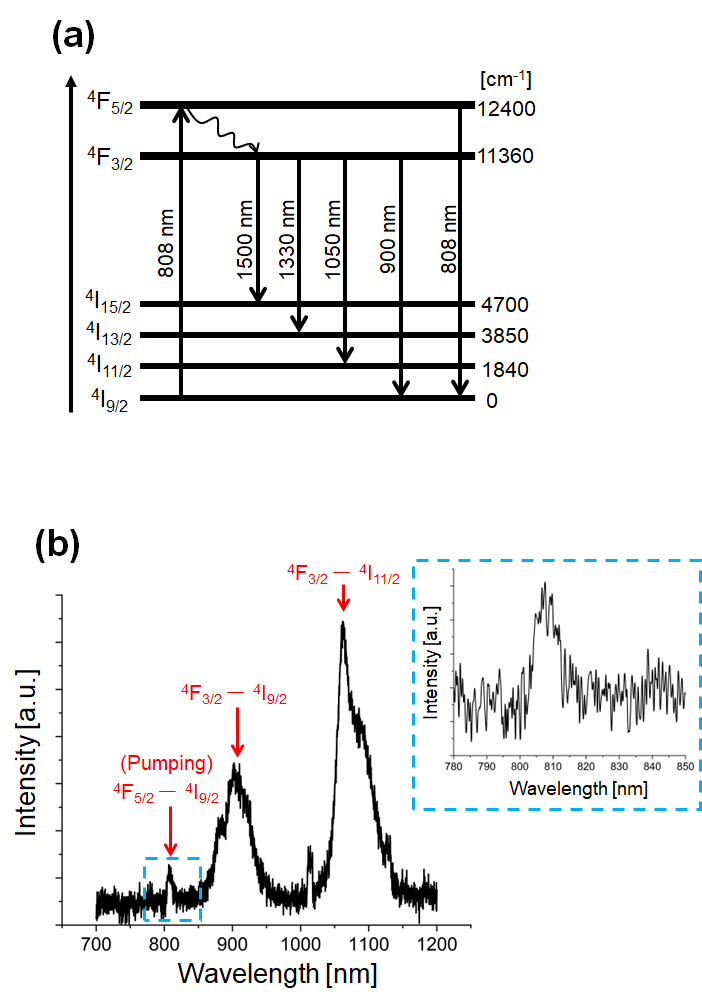}
\caption{\label{fig_concept}
(a) Energy level structure of Nd$^{3+}$ ions in silica fiber. 
(b) Emission spectrum from Nd$^{3+}$ ions doped in silica fiber before tapering used in our experiment using an $810\,\mathrm{nm}$ CW laser as a pump laser.
This spectrum was measured by using a standard Czerny-Turner-type spectrometer with Grating imaging CCD camera (HRS-300 Princeton Instrument).
}
 \end{figure}

The use of Nd$^{3+}$ ions has several advantages.
First, neodymium is one of the most abundant of the RE elements in the current industry~\cite{rare-earth03}. Therefore, it is possible to obtain the material at an affordable price under economically stable supply conditions.
Due to this availability, neodymium ions have always played important role in realizing applications in the rare-earth industry such as fiber lasers~\cite{Ndfiberlaser01} and as the dopant of RE-doped solid-state materials~\cite{Ndsolid01}. These results suggest that neodymium ions can also be a promising candidate in optical quantum information applications including single-photon sources.

Second, in terms of optical properties, neodymium has rich energy levels, as shown in Figs. \ref{fig_concept}(a) and \ref{fig_concept}(b), which enables the selection of emission wavelengths with transitions corresponding to each emission wavelength.
Compared with other RE elements, neodymium has an emission wavelength close to that of the visible region~\cite{rare-earth01}. Therefore, emitted photons from Nd$^{3+}$ ions can be sufficiently detected by silicon avalanche photodiodes, promoting the implementation of low-cost single-photon sources. 
Moreover, neodymium can generate single photons with wavelengths of 1330 and 1500 nm, as shown in Fig. \ref{fig_concept}(a), which are close to the wavelength regions of standard optical fiber telecommunications.

Third, neodymium has naturally occurring stable isotopes with nonzero nuclear spins: $^{143}$Nd($I=7/2$) and $^{145}$Nd($I=7/2$)~\cite{isotope01, isotope02}. These isotopes, which have non-zero nuclear spins, could be used in quantum memory applications with hyperfine transitions. 
Many of the experiments generating quantum entangled photon pairs by spontaneous parametric downconversion(SPDC) have been demonstrated under the same wavelength region as the neodymium emission wavelength~\cite{entanglement01, entanglement02}.
Therefore, new quantum information technologies are expected by combining single-photon sources generated by RE ions and entangled photons produced by SPDC.

 \begin{figure}[htp]
 \centering
\includegraphics[width=8cm]{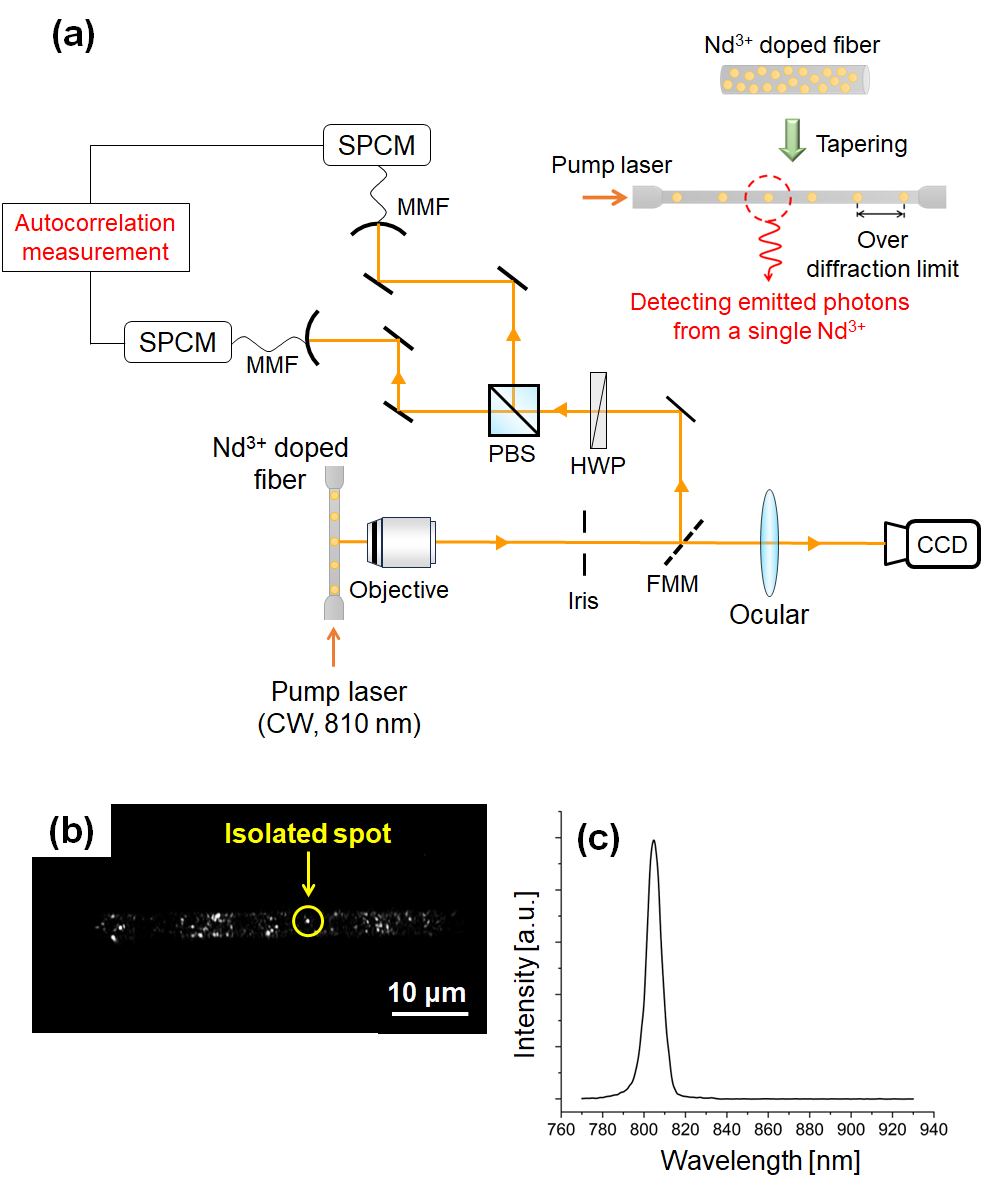}
\caption{\label{fig_setup}
(a) Experimental setup. 
A pump laser(CW, $810 \pm 2\,\mathrm{nm}$) was injected into the tapered fiber from the direction of the waveguide axis.
The images of the fiber were observed with a CCD camera using an objective lens and an ocular lens from the side of the fiber. Using an iris, an isolated single Nd$^{3+}$ ion was spatially separated.
Changing the path using flip mirror mount(FMM), photons emitted from a single Nd$^{3+}$ ion in tapered fiber sent the setup for autocorrelation measurement.
The photons were separated into two paths with a half-wave plate(HWP) and a polarizing beam splitter(PBS), and collected multimode fibers(MMFs) and single-photon counting modules(SPCMs).
(b) The images taken by a CCD camera. Our experiments focused on the emission from the Nd$^{3+}$ ion marked in the figure.
(c) Emission spectrum from a single Nd$^{3+}$ ion in tapered silica fiber observed from the side of the fiber using a monochromator and a SPCM.
}
 \end{figure}

\begin{figure}[htp]
 \centering
\includegraphics[width=8cm]{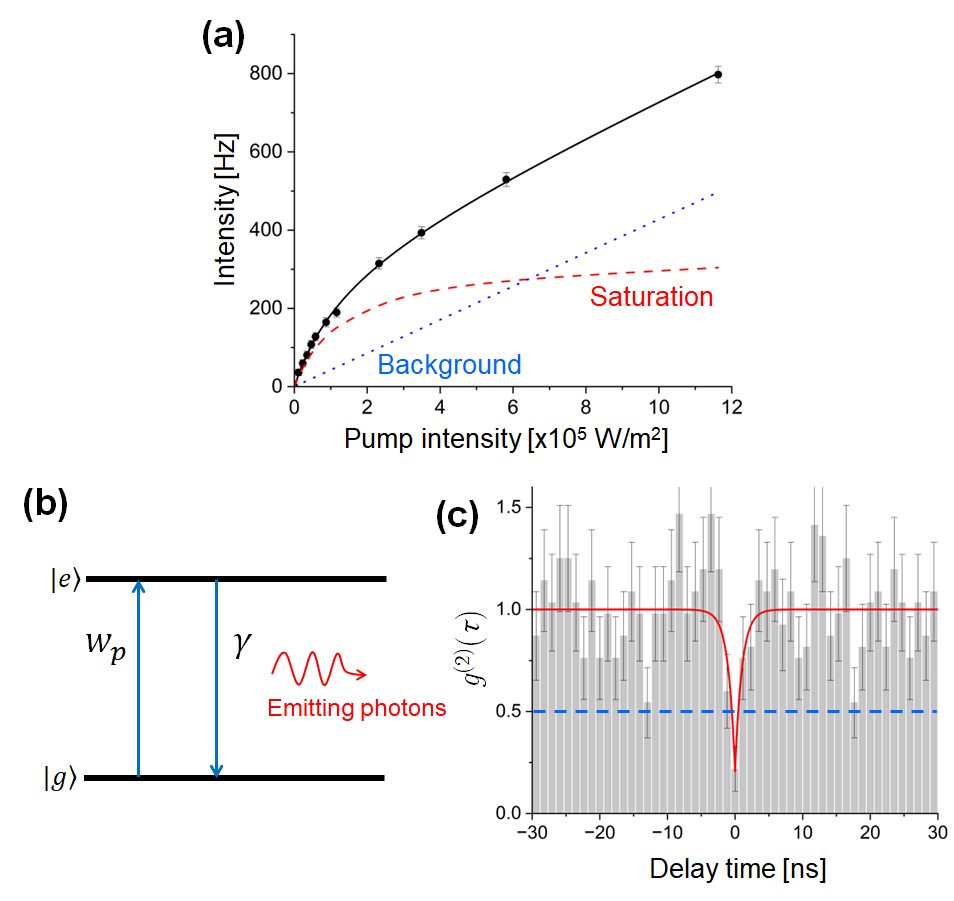}
\caption{\label{fig_result}
Experimental results. 
(a) Fluorescence intensity as a function of pump intensity with a fitting curve(black solid line). Black dots show the experimental data with the estimated error bar, and the blue dot line increasing linearly represents the background intensity due to the pump laser. The red dash curve shows the actual intensity correlation estimated by subtracting linearly increasing background from the experimentally measured intensity correlation. The value of pump intensity represents the intensity of pump light propagating in the fiber.
(b) Illustration of optical absorption and emission in a two-level system. Electrons in the ground state $\ket{g}$ are excited into the excited state $\ket{e}$ by the pump light at an absorption rate of $w_{p}$, and transition from the excited state $\ket{e}$ to the ground state at a decay rate $\gamma$ with emitting photons.
(c) Results of coincidence measurements with a fitting curve(red solid line) using a CW laser($810\,\mathrm{nm}$ wavelength) with estimated error bars. The blue dotted line is the value of the second-order correlation function, which is 0.5, and represents the boundary where the photons are really generated from a single Nd$^{3+}$ ion.}
 \end{figure}

Figure \ref{fig_setup}(a) shows our experimental setup for microscopy, spectroscopy, and measurement of saturation intensity, second-order autocorrelation function, and the dependence of polarization states.
We observed the photoluminescence from a single Nd$^{3+}$ in tapered silica fiber. This tapered fiber with single Nd$^{3+}$ ions was prepared by tapering commercially available Nd$^{3+}$ ions-doped fiber(Kokyo Inc.) with a heat and pull method~\cite{Yb01}.
The confinement condition by the fiber depends on the position of the ions into the fiber~\cite{Yb01}. Therefore, the optical transmission of the fiber is also affected by the position of the ions.
The direction of pump light to the Nd$^{3+}$ ion and collecting photons emitted from the ion is orthogonal as shown in Fig. \ref{fig_setup}(a). The excitation pump laser was injected from the input of the structured fiber and excited the ions in the tapered fiber. We observed the emitted photons from the side of the tapered fiber. In addition, the autocorrelation measurement has been done under the low excitation power that is not to saturate a Nd$^{3+}$ ion. Under the low excitation power. the signal-to-noise rate for this measurement method is about 4 estimated from Fig. \ref{fig_result}(a) at the saturation intensity. Therefore, the scattered excitation light does not effectively affect the observation of the emitted photons from the side. 

Figure \ref{fig_setup}(b) shows the images of photoluminescence from Nd$^{3+}$ ions doped in tapered fiber taken by a charged-coupled device(CCD) camera. It is found that neodymium-ions doped in tapered fiber are positioned discretely due to the tapering process.
From Fig. \ref{fig_setup}(b) with a scale bar, the tapered fiber diameter is estimated as 3.3 $\mathrm{\mu m}$.
By using an iris whose hole size is $1.2\,\mathrm{mm}$(Thorlabs, Inc., ID15/M), only one of the spots emitting photons is spatially isolated, as marked in Fig. \ref{fig_setup}(b).
In this paper, we focused on the emission for only one site marked in Fig. \ref{fig_setup}(b), and the site was separated from any other sites over the diffraction limit. Therefore, our experimental results seem to be not affected by any other sites.
The mean density of isolated single Nd$^{3+}$ ions in tapered fiber is estimated as $(1.1 \pm 0.4) \times 10^{16}\,\mathrm{m^{-3}}$. The typical toping level of rare-earth ions into the silica fiber is $10^{2} - 10^{3} \,\mathrm{ppm}$~\cite{doping01}. On the other hand, the actual number of optically active ions was affected by the intensity and transversal mode of the pump laser~\cite{Yb01}. Therefore, the estimated value of the mean density from Fig. \ref{fig_setup}(b) was not simply determined by the doping level.
In our experiment, a small, clearly blight spot was selected from the image shown in Fig. \ref{fig_setup}(b). However, it is not possible to determine whether the spot is really a single Nd$^{3+}$ ions or a cluster of multiple ions from the image only. Whether this isolated spot is actually given by a single Nd$^{3+}$ ion can be determined through autocorrelation measurement.

Figure \ref{fig_setup}(c) shows the emission spectrum from a single Nd$^{3+}$ ion in tapered fiber. The peak width of the spectrum shown in Fig. \ref{fig_setup}(c) is estimated as $8.5 \pm 0.1 \,\mathrm{nm}$. This peak width is similar to the one in Fig. \ref{fig_concept}(b) about $\sim 10\,\mathrm{nm}$.
The spectrum from the ensemble ions is easily observable under a low pump intensity as shown in Fig. \ref{fig_concept}(b). In contrast, the spectrum from a single Nd$^{3+}$ ion in the structured fiber is of quite low intensity to measure with the spectrometer. Therefore, the emission spectrum shown in Fig. \ref{fig_setup}(c) was measured by using a monochromator setup (SPG-120IR, Shimazu) with a single-photon counting module(SPCM). 
We used an excitation laser with the power much higher than the one for ensembled ions under the limited detection efficiency of SPCM. In the results, the background emission from the excitation laser becomes dominant in the spectrum, and other emission lines are hardly observable behind the large background as shown in Fig. \ref{fig_setup}(c).
In addition, it is reported that the ratio of emission peaks at each wavelength changed between fibers before and after tapering~\cite{Yb01}.
One of the factors that the ratio of emission peaks at each wavelength changed between fibers before and after the tapering is the relation between the position of an emitter in the fiber and the confinement condition by the fiber~\cite{Yb01}. Another factor is the relatively low intensity of spectrum given by the non-resonant fluorescence from a single Nd$^{3+}$ against for the one given by nearly resonant fluorescence. Therefore, the reason for the dominant peak around 810 nm is also attributed to the above-mentioned multiple factors.

Using the spatially isolated spot, we measured the dependence of the intensity of emitted photons from a single Nd$^{3+}$ ion in fiber on the pump intensity of the excitation laser, which was injected in the direction of the waveguide axis of the fiber. The experimental result is shown in Figure \ref{fig_result}(a). The saturation intensity for a single Nd$^{3+}$ ion in tapered silica fiber was estimated as $(7.1 \pm 0.8) \times 10^{4} \,\mathrm{W/m^{2}}$ corresponding to the saturation power as $0.61 \pm 0.07 \,\mathrm{\mu W}$ from the experimental saturation curve, which is fitted as follows:
  \begin{align}
I_{\mathrm{emit}} = \dfrac{F_{1}I}{2I_{\mathrm{sat.}} + I} + F_{2}I, \label{eq.sat-fit}
  \end{align}
where $I_{\mathrm{sat.}}$ is the saturation intensity, $I$ is the pump intensity of the excitation laser, $I_{\mathrm{emit}}$ is the fluorescence intensity of a single Nd$^{3+}$ ion, and $F_{1}$ and $F_{2}$ are scaling factors(see the supplementary material).
The second term in Eq. (\ref{eq.sat-fit}) represents a linear increase with background intensity caused by the pump light~\cite{nvcenter01}. Since the wavelengths of the excitation light and the emitted photons from a single Nd$^{3+}$ ion are close, it is not easy to exclude the pump laser sufficiently. 
The measurement value of the saturation power was smaller than that of other solid-state single-photon sources at room temperature, including semiconductor quantum dots~\cite{qdot01}, the center of nitrogen vacancy in diamond~\cite{nvcenter01}, and hexagonal boron nitrides~\cite{qkd01}.
This was because the saturation intensity was inversely proportional to the fluorescence lifetime($\tau_{\mathrm{life.}}$): $I_{\mathrm{sat.}} \propto \tau_{\mathrm{life.}}^{-1}$~\cite{sat.life01}.
It is possible to measure a fluorescence lifetime with a minor change of our experimental setup principally. However, it is not technically easy to measure the lifetime from a single Nd$^{3+}$ ion due to the micro-watt level low saturation intensity and the few-hundreds micro-seconds level very long fluorescence lifetime of the emitter. 
The lifetime given by the ensemble of Nd$^{3+}$ in silica fiber has been already known as about $500\,\mathrm{\mu s}$~\cite{Nd-life01}. We expected the similar level of the lifetime given by the single Nd$^{3+}$ ion.

We measured the second-order autocorrelation function $g^{(2)}(\tau)$ with the intensity of the pump laser under the saturation intensity to show the collection of emitted photons from a single Nd$^{3+}$ ion.
Figure \ref{fig_result}(b) shows the schematic of absorption and spontaneous emission in a two-level system of atoms. The second-order autocorrelation function in this system is given as follows~\cite{g2-eq01}:
  \begin{align}
g^{(2)}(\tau) = 1-\left\{ 1 - g^{(2)}(0) \right\} \exp \left[ -(w_{p}+\gamma)\tau \right], \label{eq.g2general}
  \end{align}
where $w_{p}$ is the absorption rate, $\gamma$ is the decay rate, $\tau$ is the delay time, and $g^{(2)}(0)$ is the value of the second-order autocorrelation function at zero delay. 
The measurement time window of coincidences of about 100\,ns was about a fourth-order magnitude smaller than the spontaneous radiative decay time of  Nd$^{3+}$ ion estimated as $\gamma^{-1}  \sim 500 \,\mathrm{\mu s}$~\cite{Nd-life01}. Under the measurement time areas, $w_p$ practically determines the antibunching time width, and the autocorrelation function (\ref{eq.g2general}) can be represented as $g^{(2)} (\tau) \simeq 1-\{ 1-g^{(2)} (0) \} \ \exp \left(-w_p \tau \right) $ under the measurement condition. 
The result of the coincidence measurement fitted by this equation is shown in Fig. \ref{fig_result}(c). From the fitting curve, the value of the second-order autocorrelation function at zero delay was estimated as $g^{(2)}(0) = 0.21 \pm 0.12$, and the time width of dip at zero delay was estimated as $2/w_{p}=2.0 \pm 0.9 \,\mathrm{ns}$. The measurement value of $g^{(2)}(0)$ was less than $0.5$, indicating the realization of single-photon generation using an isolated single Nd$^{3+}$ ion in tapered fiber at room temperature.
The uncertainty of $g^{(2)}(0)$ is mainly affected by uncorrelated background noise. The autocorrelation measurements were held under the condition that the intensity of the pump light was near the saturation intensity.
The error bars could be small when the autocorrelation measurements are performed under the low excitation power not to saturate a Nd$^{3+}$ ion. The ease of saturation for a single Nd$^{3+}$ ion causes an unwanted background signal for emitted photons. The emission rate of the photons also needs to be higher than the dark count rates of the detectors and background counts. Under the very low saturation power, it makes hard to satisfy both requirements simultaneously. 
In addition, the effect of time jitter cannot be negligible. We used a SPCM and time-amplitude converter(TAC) for the autocorrelation measurements. The time resolution of SPCM is 1 ns and that of TAC is 0.4 ns, which means that the time resolution of the measurement results depends on the time resolution of SPCM.
However, the uncertainty of the decay time given by the fitting curve is also comparable to the time resolution of SPCM. Therefore, we believe that the value of decay time $\sim$2 ns is reasonable within the uncertainty.
For quantum emitters with short fluorescence lifetime, we guess that the recovery time would mainly depend on the pump light intensity. On the other hand, in our experiment, the autocorrelation measurements have been done under the very low excitation power due to the easiness of the saturation of a single Nd$^{3+}$ ion. Under the condition of low pump power, the recovery time is practically determined not by the pump power but by the fluorescence lifetime of Nd$^{3+}$ ion.

 \begin{figure}[htp]
 \centering
\includegraphics[width=8cm]{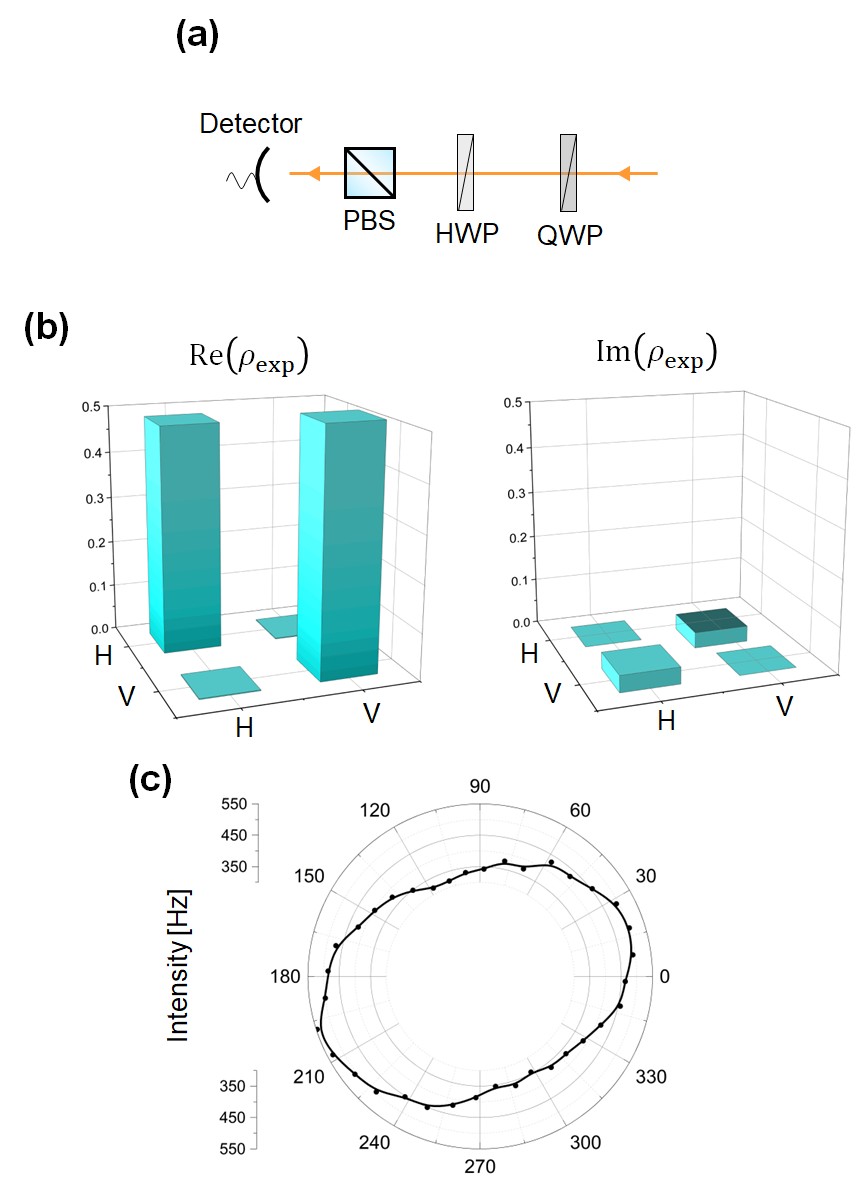}
\caption{\label{fig_polarization}
(a) Schematic of experimental setup of measurement for polarization state of photons emitted from a single Nd$^{3+}$ ions in tapered fiber.
(b) Reconstructed density matrix with real and imaginary part by quantum state tomography.
(c) Intensity of emitted photons from a single Nd$^{3+}$ in tapered fiber depending on the linear polarization states. The polar graph shows the relationship between the detected fluorescence intensity and the rotation angle of linear polarization angle as $\theta$, where $\theta /2$ is the rotation angle of HWP.
}
  \end{figure}
In order to clarify the polarization state of photons emitted from a single Nd$^{3+}$ ions in the tapered fiber, we measured Stokes' parameters and reconstructed the density matrix by quantum state tomography~\cite{polarization01}.
Figure \ref{fig_polarization}(a) shows the schematic of measurement for polarization state of photons emitted from a single Nd$^{3+}$ ions in the tapered fiber. In this measurement, a combination of a polarization beam splitter(PBS) and a half-wave plate(HWP) was used as a polarizer.
The Stokes' vector constructed by the measured Stokes' parameters is $\bm{s} \equiv (S_{1}/S_{0}, S_{2}/S_{0}, S_{3}/S_{0}) = (0.005, 0.073, -0.032)$.
The norm of $\bm{s}$ is $|\bm{s}| \sim 0$, which indicates that unpolarized single photons can be generated using an isolated single Nd$^{3+}$ ions in tapered fiber.
Figure \ref{fig_polarization}(b) shows the experimentally reconstructed density matrix($\rho_{\mathrm{exp}}$).
The fidelity($F$) of $\rho_{\mathrm{exp}}$ to the completely unpolarized state whose density matrix is represented as $\rho_{\mathrm{ideal}} = \dfrac{1}{2}
  \begin{pmatrix}
  1 & 0 \\
  0 & 1
  \end{pmatrix}
$
is calculated as $F= \mathrm{Tr}\sqrt{\sqrt{\rho_{\mathrm{ideal}}}\rho_{\mathrm{exp}}\sqrt{\rho_{\mathrm{ideal}}}} \sim 0.99$.
We also measured the intensity depending on the linear polarization states of emitted photons from a single Nd$^{3+}$ ion in the tapered fiber, and the results are shown in Fig. \ref{fig_polarization}(c).
A slight characteristic elliptical dependence for the polarization was observed. It seems that this elliptical dependence for the polarization states came from the cylindrical structure of the fiber.
One of the reasons is that the similar results have been reported observing elliptical polarization of photons from an emitter coupled to the fiber~\cite{polarization02}. This elliptical dependence was observed when the emitted photons were detected from the side of the fiber.
Another reason is that the core diameter estimated from the picture in Fig. \ref{fig_setup}(b) is about $3\,\mathrm{\mu m}$. On the other hand, the emission wavelength of Nd$^{3+}$ ions is about $1\,\mathrm{\mu m}$. Therefore, the core diameter of the fiber is about three times larger than the wavelength. The difference of the sizes is still comparable to each other within the same order of magnitude.
In addition, the confinement and resonance condition determined by the core size and the wavelength are important~\cite{Yb01}. These conditions also depend on the position of a Nd$^{3+}$ ion in the fiber. Therefore, it is expected that a different polarization state is observed depending on the position of the Nd$^{3+}$ ions in the fiber.
The detected photons were emitted from a single Nd$^{3+}$ ion through the actual absorption of pump light and the emission process through spontaneous emission.
Therefore, the polarization state of pump laser seems not to affect the polarization state of the emitted photons.
It would be beneficial to obtain data on luminescence from other sites. However in the actual experiment, about 6\,h to accumulate the data on each site were needed due to the low emission rate of Nd$^{3+}$ ions. Therefore, collecting experimental data from multiple sites is really not easy to be performed under the same experimental condition.

In conclusion, we experimentally demonstrated single-photon generation by using an isolated single Nd$^{3+}$ ion in optical tapered fiber at room temperature.
The measurement value of the second-order autocorrelation function at zero-delay was $g^{(2)}(0) = 0.21 \pm 0.12$. The value of $g^{(2)}(0)$ less than 0.5 was evidence of the generation of really single photons.
We also studied the polarization dependence of generated single photons from the emitter in tapered fiber, and observed the elliptical dependence of linear polarization states due to the cylindrical structure of the tapered fiber. 
The saturation intensity of the pump laser for a single Nd$^{3+}$ ion in tapered fiber was$(7.1 \pm 0.8) \times 10^{4} \,\mathrm{W/m^{2}}$ corresponding to the saturation power as $0.61 \pm 0.07 \,\mathrm{\mu W}$. The value of the saturation power was smaller than that of other typical solid-state emitters, which was attributed to the longer fluorescence lifetime of Nd$^{3+}$ ions in silica fiber. The introduction of cavity structures for Nd$^{3+}$-doped tapered fiber is expected to improve the efficiency of single-photon generation for practical applications.
As shown in our experimental results, the single photon emitter using Nd$^{3+}$ ions in optical fiber is expectable to realize a low-cost wavelength-selectable single-photon sources for photonic quantum applications.

\section*{Acknowledgment}
Authors would like to thank Shigeki Takeuchi and Hideaki Takashima for useful discussions. K. Shimizu has been supported by a doctoral student scholarship(2023) of Amano Institute of Technology, Japan. This work was supported by Tokyo University of Science Research Grant and JST Grant-in-Aid for Scientific Research (C) Grant No. 21K04931.

\appendix

\section{Theory of saturation intensity of two-level systems in fiber}
\begin{figure}[htp]
 \centering
\includegraphics[width=8cm]{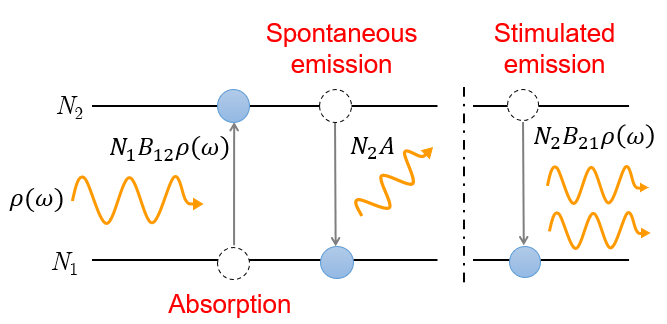}
\caption{\label{fig_app} Schematic of optical absorption and emission in a two-level system, which has the population of each level as $N_{i}\,(i=1,2)$. The atoms in this system are exited with the absorption rate represented as $N_{1}B_{12}\rho(\omega)$ because of the pump light having $\rho(\omega)$, which is the energy density per unit frequency. The excited atoms transition to a lower level with emitting photons. There are two ways to make this transition: spontaneous emission, whose transition rate is $N_{2}A$, and stimulated emission, whose transition rate is $N_{2}B_{21}\rho(\omega)$, where $A, B_{21}$, and $B_{12}$ are the Einstein coefficients for spontaneous emission, stimulated emission, and absorption, respectively.} 
 \end{figure}

The ratio equation of the system shown in Figure \ref{fig_app} is represented as
  \begin{align}
\dfrac{dN_{2}}{dt} = B_{12}\rho(\omega)N_{1} -AN_{2} - B_{21}\rho(\omega)N_{2}, \label{eq.saturation01}
  \end{align}
where $\rho(\omega)$ is the energy density of the pump light per unit frequency, $N_{i}\,(i=1,2)$ is the population of the level $i$, and $A, B_{21},$ and $B_{12}$ are the Einstein coefficients for spontaneous emission, stimulated emission, and absorption, respectively.
From the solution of Eq. (\ref{eq.saturation01}), the ratio of the population of each level in the steady state is
  \begin{align}
\dfrac{N_{2}}{N_{1}} &= \dfrac{\rho(\omega)}{A/B+\rho(\omega)} \nonumber \\
&= \dfrac{I(\omega)}{cA/B+I(\omega)}, \label{eq.saturation02}
  \end{align}
under the condition of $B_{12} = B_{21} = B$ and using the relationship between $\rho(\omega)$ and radiation intensity $I(\omega)$ represented as $I(\omega)=c\rho(\omega)$, where $c$ is the speed of light.
The power of the pump light as $P$ propagating in the fiber, which has a cross-section $S$, is represented as
  \begin{align}
P &\simeq I(\omega)\Delta \omega \times S \nonumber \\
&= \dfrac{2\pi c SI(\omega)\Delta \lambda}{\lambda^{2}}, \label{eq.saturation03} 
  \end{align}
where $\lambda$ is the wavelength of the pump light, $\Delta \omega$ is the broadening of the frequency, and $\Delta \lambda$ is the broadening of the wavelength of the pump light.
The pump light intensity at which $N_{2} / N_{1} = 1/3$ is the definition of saturation intensity. Therefore, the saturation power $P_{\mathrm{sat.}}$ of the pump light for a single Nd$^{3+}$ ion in tapered fiber is represented as
  \begin{align}
P_{\mathrm{sat.}} = \dfrac{4\pi hc^{2}S\Delta \lambda}{\lambda^{5}}, \label{eq.saturation04}
  \end{align}
using Planck's law as $A/B = 4h/\lambda^{3}$, where $h$ is Planck's constant.

For the parameters in Eq. (\ref{eq.saturation04}), the cross-section of tapered fiber shown in Figure 2(b) is estimated at about $10^{-12}\,\mathrm{m^{2}}$ and $\lambda \simeq 810\,\mathrm{nm}$ from the wavelength of the pump laser in our experiment.
The result of fitting curve in Figure 2(c), the broadening of the wavelength is estimated as $\Delta \lambda \sim 8.5\,\mathrm{nm}$
Using these parameters, the theoretical value of saturation power is calculated as $P_{\mathrm{sat.}}\simeq57\,\mathrm{\mu W}$.
However, the linewidth of the peak around 810\,nm in Figure 2(c) was broadened because our experiments used multimode lasers as the pump laser.
Therefore, the value of broadening of the wavelength used in calculating the theoretical value was estimated to be larger than the the value of broadening of the wavelength that actually contributes to the excitation.

  \end{document}